\documentclass[twocolumn]{aastex6}
\usepackage[]{units}
\usepackage[update,prepend]{epstopdf}
\usepackage{graphicx}
\usepackage{amssymb}
\usepackage{amsmath}
\usepackage{threeparttable}
\usepackage{hyperref}
\usepackage{color}

\newcommand{\oiii}{\hbox{[O$\,${\scriptsize III}]}}
\newcommand{\nv}{\hbox{N$\,${\scriptsize V}}}

\newcommand{\nii}{\hbox{[N$\,${\scriptsize II}]}}
\newcommand{\ha}{\hbox{H$\alpha$}}
\newcommand{\hb}{\hbox{H$\beta$}}

\newcommand{\kms}{km\,s$^{-1}$} 
\newcommand{\msun}{M$_{\odot}$} 
\newcommand{\eden}{cm$^{-3}$}

\newcommand{\momfluxout}{$\dot{P}_{outflow}$ }
\newcommand{\momfluxratio}{$\frac{\dot{P}_{outflow}}{\dot{P}_{AGN}}$}
\newcommand{\msigma}{$M_{\bullet}-\sigma~$}
\newcommand{\ergs}{erg s$^{-1}$ }
\newcommand{\myr}{M$_\odot$~yr$^{-1}$} 
\newcommand{\loghn}{log(\nii/\ha) }
\newcommand{\logohb}{log(\oiii/\hb) }

\mathchardef\mhyphen="2D
\shortauthors{Vayner et al.}

\begin{document}

\title{Powerful winds in high-redshift obscured and red quasars}

\author{Andrey Vayner\altaffilmark{1}, Nadia L. Zakamska\altaffilmark{1}, Rogemar A.  Riffel\altaffilmark{2,1}, Rachael Alexandroff \altaffilmark{3}, Maren Cosens\altaffilmark{4}, Fred Hamann \altaffilmark{5} Serena Perrotta\altaffilmark{4}, David S. N. Rupke \altaffilmark{6} , Thaisa Storchi Bergmann \altaffilmark{7}, Sylvain Veilleux\altaffilmark{8,9}, Greg Walth \altaffilmark{10}, Shelley Wright\altaffilmark{4},  Dominika Wylezalek\altaffilmark{11}}

\altaffiltext{1}{Department of Physics and Astronomy, Johns Hopkins University, Bloomberg Center, 3400 N. Charles St., Baltimore, MD 21218, USA}
\altaffiltext{2}{Universidade Federal de Santa Maria, CCNE, Departamento de F\'isica, 97105-900, Santa Maria, RS, Brazil} 
\altaffiltext{3}{New York City Council}
\altaffiltext{4}{Center for Astrophysics \& Space Sciences, University of California San Diego, 9500 Gilman Drive La Jolla, CA 92093 USA} 

\altaffiltext{5}{Department of Physics \& Astronomy, University of California, Riverside, CA 92507, USA}

\altaffiltext{6}{Department of Physics, Rhodes College, Memphis, TN 38112, USA}

\altaffiltext{7}{Instituto de Fisica - UFRGS
Campus do Vale, CP 15051               
91501-970 Porto Alegre - RS - Brasil}

\altaffiltext{8}{Department of Astronomy
University of Maryland,
Physical Sciences Complex,
Stadium Drive,
College Park, MD 20742-2421}

\altaffiltext{9}{Joint Space-Science Institute, University of Maryland, College Park, MD 20742}

\altaffiltext{10}{Observatories of the Carnegie Institution for Science,
813 Santa Barbara Street,
Pasadena, CA 91101
USA}

\altaffiltext{11}{Zentrum für Astronomie der Universität Heidelberg
Astronomisches Rechen-Institut
Mönchhofstr, 12-14 
69120 Heidelberg 
Germany}

\begin{abstract}
Quasar-driven outflows must have made their most significant impact on galaxy formation during the epoch when massive galaxies were forming most rapidly. To study the impact of quasar feedback we conducted rest-frame optical integral field spectrograph (IFS) observations of three extremely red quasars (ERQs) and one type-2 quasar at $z=2-3$, obtained with the NIFS and OSIRIS instruments at the Gemini North and W. M. Keck Observatory with the assistance of laser-guided adaptive optics. We use the kinematics and morphologies of the \oiii\ 5007\AA\ and \ha\ 6563\AA\ emission lines redshifted into the near-infrared to gauge the extents, kinetic energies and momentum fluxes of the ionized outflows in the quasars host galaxies. For the ERQs, the galactic-scale outflows are likely driven by radiation pressure in a high column density environment or due to an adiabatic shock. The outflows in the ERQs carry a significant amount of energy ranging from 0.05-5$\%$ of the quasar's bolometric luminosity, powerful enough to have a significant impact on the quasar host galaxies. The outflows are resolved on kpc scales, the observed outflow sizes are generally smaller than other ionized outflows observed at high redshift. The high ratio between the momentum flux of the ionized outflow and the photon momentum flux from the quasar accretion disk and high nuclear obscuration makes these ERQs great candidates for transitional objects where the outflows are likely responsible for clearing material in the inner regions of each galaxy, unveiling the quasar accretion disk at optical wavelengths.

\end{abstract}


\section{Introduction}
\label{sec:intro}

Detecting winds driven by active galactic nuclei (AGN) and identifying mechanisms of AGN feedback is at the forefront of observational astronomy because of the critical importance of these processes in galaxy formation \citep{silk98, king03}. This task is made difficult by the multi-phase nature of galactic winds, where different phases of the wind must be probed using different observational techniques \citep{Veilleux20}. At relatively low redshifts ($z<1$), AGN-driven outflows have been detected and studied through nebular emission lines \citep{morg05,holt08,rupk11,liu13a,harr14} as well as molecular and neutral gas phases \citep{cico14,alat11,aalt12,feru14,morg13,Strum11,Veilleux17,Fluetsch19,Lutz20}.

At low redshifts, the massive stellar component of the host galaxies is already in place. The dramatic decline of the galaxy stellar mass function beyond $M^*$ is likely due to AGN feedback \citep{crot06, hopk06} and must have been established at earlier epochs. To probe the effects of AGN on the formation of the massive galaxies, we must examine the period $z=2-3$ \citep{boyl98,mada14} when galaxies were growing most rapidly.

Increasing body of work shows that high-redshift quasars often display forbidden emission lines (such as \oiii $\lambda$5007\AA) with high velocity offsets or dispersions, which may be positively correlated with quasar luminosity, obscuration and level of jet activity \citep{zaka16b, wu18,glik17,urru12,brus16,pern15a,Forster18,Forster19,Davies19,Kakkad20}. Although the nominal velocities of \oiii\ are often well in excess of 1000 \kms, the impact of these ionized gas outflows on their hosts cannot be established unless there are spatially resolved observations. Spatial information is necessary to demonstrate galaxy-wide extents, high velocities and outflow rates with significant momentum flux on galactic and circumgalactic scales or direct impact on the molecular gas reservoir. Such observations exist for some sub-samples \citep{nesv06, nesv08, cano12,Swinbank15,carn15, kakk16,Harrison16,vayner17,Brusa18,Herrera-Camus19,Davies20}. However, this task remains extremely challenging due to the cosmological surface brightness dimming and due to the difficulty of subtracting the quasar light to reveal the faint signatures of the wind.

Despite the significant ongoing observational effort, it is not clear how the kinetic power of the outflows depends on the quasar evolutionary stage or type of activity, or how the ionized gas outflows are related to other phases. Supermassive black boles (SMBH) grow the majority of their mass in obscured environments \citep{hopkins05}. Once they reach a critical mass theoretical work predicts that the central accretion disk can produce outflows powerful enough to extend onto galactic scales \citep{king15}, affecting the star-forming properties of the host galaxy and clearing the central nuclear region of gas and dust. The most powerful outflows are most efficiently driven in obscured environments, either through radiation pressure or through adiabatic shocks because it is easier to couple the momentum of both the photons and the shocks to the surrounding gas \citep{wagn12,fauc12a,costa18}. The transitional phase is very rapid occurring on million-year time scales \citep{hopkins05} in numerical simulations. Hence catching AGN during this key phase can be very difficult.

Studying obscured AGN with evidence for powerful outflows can provide clues as to how they affect the star-forming properties of galaxies and how AGN transition to an unobscured phase. Extremely red quasars \citep{ross15,hama17} were initially discovered through a selection based on high rest frame ultra-violet to infrared colors (i-W3$>4.6$ mag) from the Wide-Field Infrared Survey Explorer (WISE; \citealt{wrig10}) and SDSS. Recent X-ray observations for 11 ERQs reveals high nuclear obscuration with column densities up to $10^{24}\rm~cm^{-2}$ \citep{goul18a,Ishikawa21}. ERQs show peculiar rest-frame UV emission lines with large equivalent widths, missing wings in the broad emission lines, and unusual emission-line ratios typically not seen in other quasar populations. Powerful outflows driven in extremely obscured environments may link to the atypical emission-line properties in ERQs \citep{hama17,alex18}.

While there have been excellent studies focusing on ionized outflows in systems with obscured nuclei \citep{pern15a,pern15b,cres15}, there have yet to be studies of the unique ERQ population with IFS observation. ERQs are some of the most intrinsically luminous obscured objects at high redshift. Among the obscured population at high redshift, ERQs show some of the most powerful outflows, as evidenced by extremely broad \oiii\ emission lines with velocities reaching up to 6700 \kms, not seen in any other quasar population at any redshift \citep{perr19}. Obscured AGN may be the clue to the short-lived transitional stage where feedback that affects star formation may be visible. In this paper, we present integral-field unit observations of five powerful red and obscured quasars at $z=2.3-3$ and map their ionized gas winds. In Section \ref{sec:data} we describe the sample selection, observations and data reduction. In Section \ref{sec:analysis}, we present the kinematic analysis and ionized gas maps, with individual objects analyzed in-depth in Section \ref{sec:individual}. We discuss the implications of our findings in Section \ref{sec:disc} and conclude in Section \ref{sec:conc}. We use $h=0.7$, $\Omega_{\rm m}=0.3$, $\Omega_{\Lambda}=0.7$ cosmology and air wavelengths for the emission lines, all magnitudes are on the AB system. 

\section{Observations and data reduction}
\label{sec:data}

\subsection{Sample selection}

The targets in this study are selected from two different, but physically related samples. One sample is that by \citet{alex13}, who identified 145 high-redshift obscured (type-2) quasar candidates from the spectroscopic database of the Sloan Digital Sky Survey (SDSS; \citealt{daws13}) Data Release 9 \citep{ahn12} based on emission-line widths, requiring full width at half maximum (FWHM)$<$2000 \kms\ for both the CIV$\lambda$1550\AA\ and Ly$\alpha$ emission lines. The lack of broad-emission components in the permitted lines of quasar spectra is usually taken to be a sign of nuclear obscuration, which prevents the observer from seeing the broad-line region. Therefore, one classical selection of type-2 AGN relied on the dominance of narrow lines at optical wavelengths \citep{hao05a, reye08}. Because of the greater opacity of the dust at UV wavelengths, the line-based selection of type-2 candidates at rest-frame UV wavelengths results in a sample with a wide range of obscuration values, and some of these objects are not obscured enough to be classified as type-2 based on their rest-frame optical spectra \citep{gree14b}. Indeed the type-2 quasar in this study shows a broad \ha\ emission line and is likely better characterized as a type 1.5 or 1.9 quasar. However, throughout the paper, we will be referring to this object as a type-2 quasar. The other targets in this study are selected from the ERQ sample.

The ERQ and the type-2 samples represent a large fraction of all known obscured, luminous, high-redshift quasars, they have been extensively studied across the electromagnetic spectrum by us and by other groups. In particular, we conducted extensive near-infrared (NIR) observing campaigns to measure integrated \oiii\ properties of the objects in both samples. The rest-frame optical properties of the line-selected type-2 quasar candidates are described by \citet{gree14b}, and two additional NIR long-slit spectra were published by \citet{alex18} and \citet{zaka19}. All of these observations have been done in seeing-limited conditions and are spatially unresolved. NIR (rest-frame optical) properties for ERQs are summarized by \citet{zaka16b} and \citet{perr19}. ERQs stand out from the type-2 quasars and the rest of the AGN population in that they show very broad (FWHM $\sim 1000-6000$ \kms) forbidden emission lines. These large velocity widths indicate very fast outflows, with possibly a very large opening angle. Their fast velocities translate to large outflow momenta flux and outflow kinetic luminosities, making them one of the most powerful ionized outflows among any AGN sample in the distant Universe \citep{perr19}.

From the sub-samples with available NIR spectroscopy, the targets for Near-Infrared Integral Field Spectrograph (NIFS) observations were then selected based on two criteria. The most constraining criterion is the availability of relatively bright nearby stars which can enable laser-guided adaptive optics (AO) observations; specifically, a $r<17$ mag star within 25\arcsec\ of the science target is typically required. This cut renders over 90\% of targets inaccessible to AO-assisted observations at the Gemini North Observatory. Secondly, based on the measurements from the long-slit NIR data we prioritized the most \oiii-luminous targets for our integral-field-unit (IFU) observations. These requirements, plus long requisite exposure times, limited our NIFS proposals to only one-two targets at a time. SDSS J2323$-$0100 was targeted for OSIRIS follow up after its VLT long-slit spectrum was published by \citet{zaka16b}. 

The observations presented here were built up over many semesters of proposals and observations, and the program was significantly delayed by the mechanical problems with the Gemini AO system in 2017-2018. The targets are listed in Table \ref{obstab}, with their `type' designated as type-2 or ERQ depending on their original selection as described above, and listing program numbers for the programs that were successful in obtaining relevant data.

\subsection{Observations}

NIFS on Gemini-North provides spectroscopy with $R\sim 5000$ over a 3$^{''}\times$3$^{''}$ field of view \citep{mcgregor03} using an imaging slicer design. The priority for our program was to observe the \oiii+H$\beta$ emission-line complex. Depending on the redshift of each target, this complex falls in the $H$ or $K$-band. All sources whose redshift places these lines in the regions of low atmospheric transparency were excluded at an early stage in our sample selection. For two sources with \oiii+H$\beta$ in the $H$-band, we additionally have $K$-band observations that cover the [NII]+H$\alpha$ line complex. Observations are obtained in each band separately using the same $H-K$ filter, but different gratings with central wavelengths adjusted to cover the lines of interest optimally. 

Observations were requested and conducted in the queue mode. The requested weather conditions are determined by the performance requirements of the AO system (image quality in the top 70\% of all observations; cloud cover in the top 50\% of all observations; any sky brightness and water vapor). During observations, the seeing as measured using CFHT's skyprobe are listed in Table \ref{obstab}. To enable sky subtraction, we nod by 30'' from the science target to blank sky in the repeating ``object-sky-object'' sequences. The length of exposure is a trade-off between minimizing the read noise and the read-out time, avoiding saturation by the thermal background, and obtaining enough redundancy between exposures. Individual exposure times varied between 450 and 645 sec, and total on-target exposure times varied between 20 and 100 min.

OSIRIS on Keck provides spectroscopy with $R\sim 3500$ with a variable field-of-view (FOV) utilizing a lenslet design \citep{Larkin06,Mieda14,Boehle16}. The OSIRIS observations were conducted in the Hn3 and Kbb filters using the 50 mas plate scale mode, providing a field of view of 2.4\arcsec$\times$3.2\arcsec and 0.8\arcsec$\times$3.2\arcsec, respectively. The $H$ band observations were targeted to search for extended emission on arcsecond scales, hence the larger FOV was selected, and we observed a dedicated sky frame. We optimized the $K$-band observations for on-source exposure time by nodding the object in the long axis of the OSIRIS FOV and subtracting pair observations for sky emission removal. We bracketed the $K$-band observations with observations of a PSF star selected to have the same offset from the tip/tilt star as the quasar.

\begin{table*}
	\centering
	\caption{Observing log.}
	\label{obstab}
	\begin{tabular}{cccccccc} 
		\hline
		Source & Project & Date & Instrument & Band & Exp. Time (s) & Seeing & Corrected PSF \\
		SDSS   &         &                   &                      &                        \\
		
		\hline
		 &  &  & Type-2 &  &  & \\
		 \hline
			    J081257.15+181916.8 &  GN-2014A-Q-42 & 2014/04/11& NIFS  & K & 6$\times$645 & 0\farcs37$\pm$0\farcs04 & 0\farcs17$\times$0\farcs22 \\
	                  & GN-2014B-Q-44 & 2014/11/09& NIFS & H & 5$\times$645  &
	   0\farcs49$\pm$0\farcs05 &  0\farcs31$\times$0\farcs32 \\
	   	\hline
		 &  &  & ERQ &  &  & \\
		 \hline
		J165202.64+172852.3 & GN-2019A-FT-111 &2019/05/07-08 & NIFS & K & 10$\times$600 & 0\farcs41$\pm$0\farcs09 & 0\farcs17$\times$0\farcs21 \\
	    J082653.42+054247.3 & GN-2016A-FT-8 & 2016/02/27 & NIFS     & H & 12$\times$450 & 0\farcs35$\pm$0\farcs08 & 0\farcs29$\times$0\farcs31 \\
	   J091301.33+034207.6 \tablenotemark{a} & GN-2014B-Q-44 & 2015/01/28& NIFS & K & 2$\times$645 &
	   0\farcs32$\pm$0\farcs06 & -- \\
	   J232326.17-010033.1 & & 2017/08/13 & OSIRIS & Hn4 & 10 $\times$600 & 0\farcs50  & 0\farcs085$\times$0\farcs088 \\
	   J232326.17-010033.1 & & 2017/09/02 & OSIRIS & Kbb & 6 $\times$600 & 0\farcs60  & 0\farcs082$\times$0\farcs08\\
		\hline
	\end{tabular}
	\tablenotetext{a}{Data quality not sufficient to perform any analysis.}
\end{table*}

\subsection{Data reduction}

We performed the NIFS data reduction using the  {\sc gemini iraf/pyraf} package, following the standard procedure for NIR spectra. This procedure includes the trimming of the images, flat-fielding, cosmic ray rejection, sky subtraction, wavelength, and s-distortion calibrations. To remove the telluric absorption features, we use the spectra of A0-A2V standard stars observed before or after the observation of the science target. We use the {\sc nftelluric} task of the {\sc nifs.gemini.iraf} package to normalize and shift the spectrum of the standard star and divide it into the science frames. For each frame, we use the telluric star spectrum that was observed with the most similar air mass and closest in time to the science target.

Then, we performed the flux calibration by interpolating a black body function to the spectra of the telluric standard stars and generate datacubes for each individual exposure with 0\farcs05 width square spaxels. Finally, we median combine the individual datacubes into a single datacube for each target using the {\it gemcombine} task of the {\sc gemini iraf} package and 
the {\it sigclip} algorithm to eliminate the remaining cosmic rays and bad pixels, using the peak of the continuum as reference for the astrometry of the distinct cubes.

We reduced the OSIRIS observations using the OSIRIS data reduction pipeline \citep{OSIRIS_DRP}. The pipeline first creates a master dark by median combining several dark observations taken at the start of the night. After dark subtraction, the pipeline extracts the spectra using a Lucy-Richardson deconvolution using a known spectral-PSF for each lenslet stored in the rectification matrix. The pipeline then applies wavelength calibration and combines the spectra into a 3D data cube. The pipeline performs scaled sky subtraction by scaling families of OH emission lines in the sky cubes to match the science observations. The cubes are finally mosaicked using a sigma clipping algorithm, and the data are flux calibrated similarly to the NIFS observations. Our estimated absolute flux uncertainty is at 10-15$\%$.

\subsection{Point-spread function analysis}

Quasars often outshine the extended emission from their host galaxy. Observing type-2 objects often helps with the contrast, however, at higher redshifts, the unresolved component from the narrow line region can still be brighter than that of the extended emission from the host galaxy. Also, UV-selected type-2 quasars can still show emission from the broad-line region in the Balmer series \citep{gree12} and continuum emission from the accretion disk. Our sample contains one UV-selected type-2 quasar and 3 ERQs. These sources have varying levels of contrast between the unresolved and extended emission. All sources show broad-line emission that is expected to outshine or be comparable to the magnitude of the host galaxy, hence for each object, we performed PSF subtraction before searching for extended emission from the host galaxy.

Details of the PSF subtraction routine can be found in \cite{vayner16}, and \cite{vayner20}. We utilize channels containing emissions from the quasar continuum or the broad-line region that do not overlap with OH emission from the night-sky to construct a PSF image. The image is then normalized to the peak emission and subtracted from the rest of the data channels. This technique effectively removes any unresolved emission from the data cube and leaves behind only extended emission.

\section{Analysis}
\label{sec:analysis}

\subsection{Fitting emission lines}

After the data cubes are PSF-subtracted, we search all the channels for extended emission in each emission line. For any detected emission 3$\sigma$ above the background, we collapse the data cube around the peak of the emission line and divide by the standard deviation in an empty sky region to construct a signal-to-noise (SNR) image. Spaxels with a 2$\sigma$ or above emission are then fit with a Gaussian model plus a zero degree polynomial using the least-squares fitting routine \textit{curvefit} within \textit{SciPy}. We visually inspect each fit. In some spaxels, a multi-Gaussian fit is necessary for individual emission lines. We deem a multi-Gaussian fit successful if we find an improvement in the reduced $\chi^{2}$ value. In all cases, the $\chi^{2}$ value improved by at least 2.7, after we introduced a second Gaussian component to the fit. We visually inspected the new fit and the residuals to confirm the results. 

When fitting the \ha\ emission line all parameters (intensity, width and redshift) are free. When fitting the \nii\ emission lines, we fix the redshift and the line widths to those of \ha. The ratios of the line peaks for the \nii\ 6548 \AA\ and 6583 \AA\ emission lines is held fixed at 1:2.95. The only free parameter is the line intensity of the rest-frame 6584 \AA\ \nii\ emission. When fitting the \oiii\ emission lines, the \oiii 4959 \AA\ has no free parameters, since we hold the line width and redshift fixed relative to \oiii\ 5007 \AA\, and we hold the ratio between the two line peaks fixed at 1:2.98. Channels with strong OH emission are weighted during least-squares fitting using a noise spectrum constructed from an empty sky region.

\begin{figure*}[!th]
    \centering
    \includegraphics[trim=5cm 6cm 5cm 6cm, clip=true, width=\linewidth]{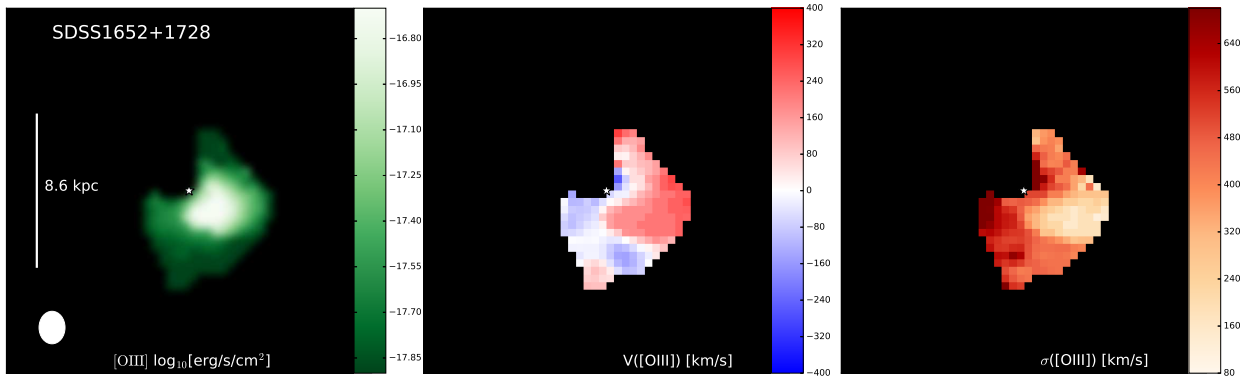}
    \caption{From left to right, PSF-subtracted \oiii emission line flux, mean velocity and velocity dispersion of SDSS~J1652+1728 over the field of view of NIFS. North is up, east is to the left. The star represents the location of the subtracted out quasar. The bar represents 1\arcsec or 8.6 kpc at the redshift of the source. The ellipse in the lower left corner shows the FWHM of the AO corrected PSF.} 
    \label{fig:1652}
\end{figure*}
\begin{figure*}[!th]
    \centering
    \includegraphics[trim=5cm 6cm 5cm 6cm, clip=true, width=\linewidth]{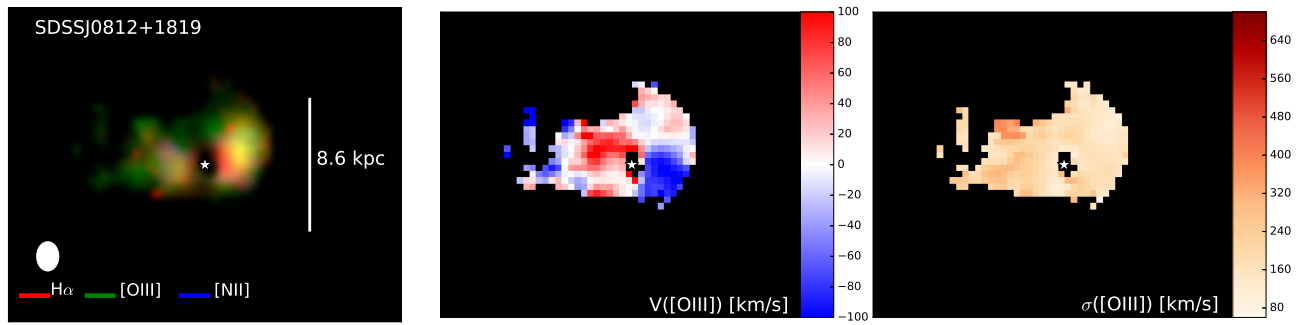}
    \caption{PSF-subtracted NIFS observations of SDSS~J0812+1819. On the left we present a three-color composite of ionized emission from \ha (red) \oiii (green) and \nii (blue) in the host galaxy of SDSSJ0812+1819. Middle panel shows the radial velocity relative to the redshift of the quasar and the right panel shows the velocity dispersion. White bar represents one arcsecond on sky and the ellipse represents the FWHM of the PSF of the LGS-assisted observations} 
    \label{fig:0812}
\end{figure*}
\begin{figure*}[!th]
    \centering
    \includegraphics[width=8.0 in]{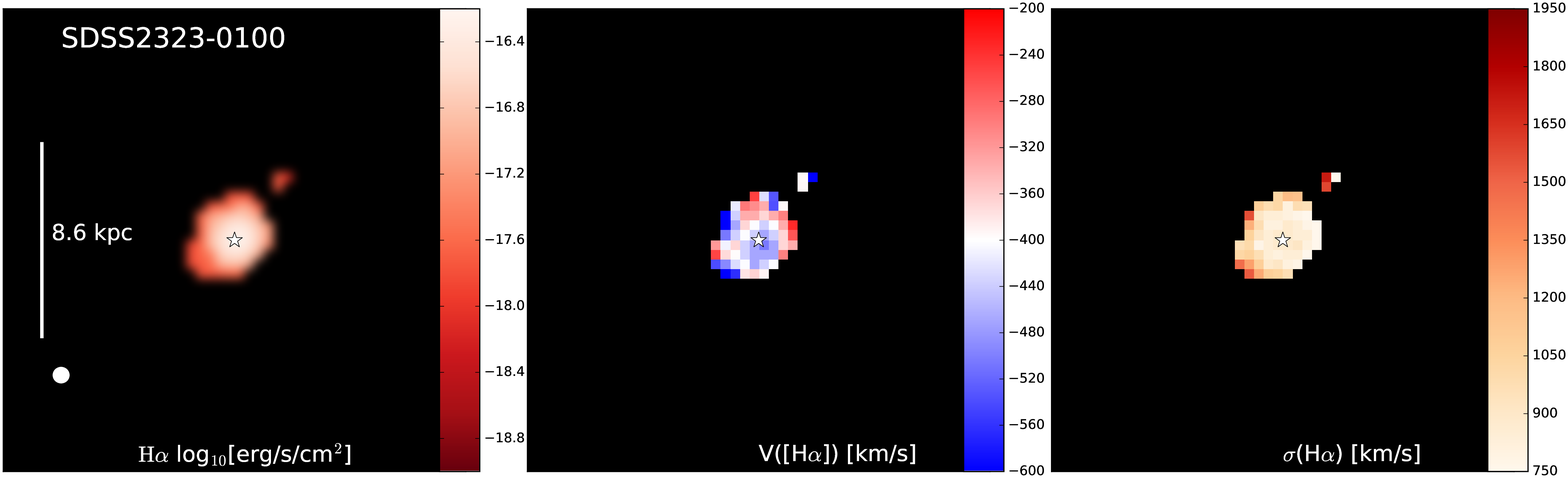}
    \caption{From left to right, PSF-subtracted \ha\ emission line flux, mean velocity, and velocity dispersion of SDSS~J12323 over the field of view of OSIRIS. North is up, east is to the left. The bar on the left figure represents one arcsecond or 8.6 kpc at the redshift of the quasar, while the ellipse represents the FWHM of the PSF in our observations.}
    \label{fig:SDSS2323}
\end{figure*}

We constructed an intensity map by summing up each emission line spectrally from -3$\sigma_{pixels}$ to +3$\sigma_{pixels}$, where the standard deviation comes from the fit of the Gaussian model. We constructed a radial velocity map by computing the Doppler shift of the line-centroid relative to the quasar's redshift. The redshifts are taken from the narrowest nebular (e.g., \hb, \oiii) emission line from the spatially-unresolved point source component. The line centroid comes from the Gaussian fit. For spaxels with a multi-Gaussian fit, we use the line-intensity weighted centroid. We constructed a velocity dispersion map using the best fit $\sigma$ value. In the case of a multi-Gaussian fit, we used the line width, where we integrated 66$\%$ of the total line intensity. For the three objects which show extended emission we present the emission line, radial velocity and dispersion maps in Figures \ref{fig:1652}, \ref{fig:0812} , and \ref{fig:SDSS2323}. OSIRIS observations of SDSSJ2323-0100 cover \oiii\ 5007 \AA\, but the wavelength coverage is too narrow to deblend \oiii\ 4959 \AA, \oiii\ 5007 \AA\ and \hb\ for us to be able to construct the maps of the \oiii\ emission line, the goal of those observations was to gauge the extent of the \oiii\ emitting gas.

\pagebreak

\subsection{Searching for outflows}
\label{sec:outflow-rates-energetics}

The most turbulent discs in massive ($>10^{11}$\msun) galaxies show an ionized gas velocity dispersion of 250 \kms\ in distinct star-forming regions with maximum rotational velocity of 400 \kms\ \citep{Forster18}. Estimates of the stellar mass for ERQs from recent HST observations \citep{zaka19} overlap with the most massive galaxies in the SINS survey \citep{Forster18}. Hence emission lines that have a radial velocity dispersion $>$250 \kms\ or radial velocity $>$400 \kms\ (computed from the line-centroid) results in gas that most likely belong to an outflow. Both the radial velocity dispersion and offset are relative to the redshift of the quasar. Spaxels satisfying these criteria have their spectra extracted to construct a high signal-to-noise ratio spectrum. The spectra are fit with multiple Gaussians for each nebular emission line, which are detected at a 3$\sigma$ or greater significance. We also search for unresolved-nuclear outflows in our systems by subtracting a model of the extended emission and extract the spectrum from the unresolved point source. The model that we subtract is a data cube consisting of the best fit Gaussian emission at each spaxel location. Similarly, we search for broad and highly kinematically offset emission-line gas.

We compute the outflow rates following the methodology presented in \citet{cano12}, assuming a conical outflow where the gas fills a significant fraction of the cone or a conical outflow consisting of many concentric shells. The reason for this assumption is that we see outflowing gas at a large range of velocities ($\pm2000$ \kms in the case of SDSSJ1652+1728, SDSSJ2323-0100 and SDSSJ0826+0542), likely due to projection effects of an inclined cone projected onto the sky. The following formula gives the mass outflow rate:

\begin{equation}\label{equation:outflow-cone}
    \dot{M}=3\frac{Mv_{out}}{R_{out}},
\end{equation}
where $M$ is the mass of the ionized gas, $v_{out}$ is the velocity, and $R_{out}$ is the extent of the outflow. For the velocity we use the non-parametric $v_{10}$ value of the \oiii\ or \ha\ emission lines; this is the velocity where 10\% of the line flux is integrated for the Gaussian component associated with the outflow. The velocity is calculated from  the spectrum integrated over the entire outflow region in each source. For the radius, we use the extent where we integrated 90\% of the flux associated with broad extended emission using the emission line map from the broad Gaussian component. For unresolved outflows, $R$ is the FWHM of the PSF. Our methodology for calculating the radius and the outflow velocity is selected to account for inclination effects of the outflow \citep{gree12,cano12}. To compute the mass, we use the luminosity of the \ha\ or \oiii\ emission line and convert that to an ionized gas mass using the methodology presented in \citet{oste06} or \citet{cano12}. For the case of \oiii\, the ionized gas mass is less certain as we do not know the metallicity and assume solar value. We also do not apply any ionization correction and assume all of the oxygen is double ionized. The contribution from other ionization states of Oxygen is unknown, and this is likely a significant contributor to additional uncertainty in deriving the outflow rate from the \oiii\ line. It further adds to the discrepancy between the measurement of outflow rates and energetics from the \oiii\ line and the Balmer lines \citep{demp18}. The largest uncertainty comes from the unknown electron density of the gas. We use a nominal value of 500 \eden\ with a range of 100-900 \eden. These values are all within the range of measured electron densities in ionized outflows at high redshift \citep{harr14,vayner17,Forster19}.

In addition to the outflow rates we also compute the momentum flux of the outflow: 

\begin{equation}
    \dot{P} = \dot{M} \times v_{out}
\end{equation}

and the kinetic luminosity:

\begin{equation}
    \dot{E} = \frac{1}{2}\times\dot{M}\times v_{out}^{2}.
\end{equation}

The radius, velocity, outflow rates and energetics of individual outflows along with the bolometric luminosity of the quasars are presented in Table \ref{tab:outflow-prop}. The bolometric luminosities are taken from \cite{gree12} and \cite{perr19}. The uncertainties on the outflows and energetics are dominated by systematic rather than measurement errors that are presented in Table \ref{tab:outflow-prop}.

\begin{deluxetable*}{llllllllll}

\tablecaption{Outflow properties of the sources within this study \label{tab:outflow-prop}. R$\rm_{out}$ is the radial extent of the outflow. V$\rm_{out}$ is the velocity of the outflow. dM/dt$\rm_{[OIII]}$ and dM/dt$\rm_{H\alpha}$ are the outflow rate using masses derived from their respective line luminosities. $\dot{P}$ is the momentum flux of the outflow. L$\rm_{AGN}$ is the bolometric luminosity of the quasar. $\frac{\dot{E}_{outflow}}{\dot{L}_{AGN}}$ is the energy coupling efficiency between the kinetic luminosity of the outflow and bolometric luminosity of the quasar. $\frac{\dot{P}_{outflow}}{\dot{P}_{AGN}}$ is the momentum flux ratio between the ionized outflow and the photon momentum flux from the quasar accretion disk. N = Unresolved nuclear outflow, E = resolved, extended outflow.}

\tablehead{\colhead{Source}&
\colhead{R$\rm_{out}$}&
\colhead{V$\rm_{out}$}&
\colhead{dM/dt$\rm_{[OIII]}$}&
\colhead{dM/dt$\rm_{H\alpha}$}&
\colhead{$\dot{P}$}&
\colhead{$\dot{E}$}&
\colhead{L$\rm_{AGN}$}&
\colhead{$\frac{\dot{E}_{outflow}}{\dot{L}_{AGN}}$}&
\colhead{$\frac{\dot{P}_{outflow}}{\dot{P}_{AGN}}$}\\
\colhead{SDSSJ}&
\colhead{kpc}&
\colhead{\kms}&
\colhead{\myr}&
\colhead{\myr}&
\colhead{$10^{35}$dyne}&
\colhead{$10^{43}$ \ergs}&
\colhead{$10^{47}$ \ergs}&
\colhead{$\%$}&
\colhead{}
}
\startdata
0812N  &$<$1.9         & 650$\pm$10  & $>$35$\pm$18 & $>$62$\pm$45 &  $>$2.5$\pm$1.8 & $>$0.8$\pm$0.6 & 0.5$\pm$0.1 & $>$0.02$\pm$0.01 & $>$0.15$\pm$0.1\\
0826N  & $<$2.55       & 2829$\pm$10 & $>$1340$\pm$680 & -- &  $>$240$\pm$120 & $>$338$\pm$170 & 3$\pm0.3$ & $>$1$\pm$0.5 & $>$2$\pm$1\\
1652N  & $<$1.7        & 1334$\pm$10 & $>$843$\pm$430 & -- &  $>$70$\pm$40 & $>$47$\pm$20 & 5$\pm$0.1 & $>0.1\pm0.05$ & $>$0.4$\pm$0.2\\
1652E  & 2.5$\pm$0.1   & 1031$\pm$50 & 124$\pm$100 & -- &  8$\pm$7 & 4$\pm$3 & 5$\pm$0.1 & 0.01$\pm$0.006 & 0.05$\pm$0.04\\
2323E   & 0.77$\pm$0.01 & 1963$\pm$50 & -- & 2750$\pm$2465 &  340$\pm$305 & 334$\pm$300 & $1\pm0.1$ & $2.7\pm2.5$ & 8.5$\pm$8\\
\enddata
\end{deluxetable*}

\section{Individual Objects}\label{sec:individual}
\pagebreak
In this section, we briefly describe previously known properties for each object and present their new NIR IFU data. 

\subsection{SDSS~J1652+1728 (z=2.9482)}
\label{sec:1652}
SDSS~J1652+1728 was part of the ERQ core sample \citep{ross15, hama17}, but its rest-frame UV line widths are just above the 2000 \kms\ required to satisfy the type-2 quasar candidate criteria from \citet{alex13}. GNIRS spectrum of this object which showed an extremely broad blue-shifted component in its \oiii\ line \citep{alex18}, quite in line with other ERQs which show kinematically extreme \oiii\ \citep{zaka16b, perr19}. The object is a $\sim 1.5$ mJy radio source, which lies a few $\sigma$ away from the relationship between \oiii\ velocity width and radio power, in the sense that its radio luminosity is perhaps above that expected as a by-product of the outflow alone \citep{hwan18}. 

Spectropolarimetric observations \citep{alex18} of SDSS~J1652+1728 -- a unique probe of circumnuclear geometry of the outflow -- show a 90$^{\circ}$ swing in the polarization position angle for rest-frame UV emission lines (L$\alpha$, CIV and NV) as a function of velocity indicating that the outflowing material originates within an equatorial dusty scattering region. Furthermore, SDSS~J1652+1728 is the most highly polarized source in the sample, with continuum polarization reaching 10$-$15\% and polarization position angle 130$^{\circ}$ East of North. This is relevant because, from this value, we can infer the dominant orientation of the scatterer relative to the nucleus, 220$^{\circ}$ East of North. X-ray observations \citep{goul18a,Ishikawa21} show that SDSS~1652+1728 is a Compton thick, intrinsically luminous quasar with a column density of 1$\times10^{24}~\rm cm^{-2}$ and X-ray luminosity $L_{2-10~keV}=1.27\times10^{45}$ \ergs.

HST observations \citep{zaka19} in the rest-frame $B$-band show an extended host galaxy with multiple clumps or companions and an extended low surface brightness feature directed to the West, which may be a tidal tail. The host luminosity is well above the median for the ERQs and is $\ga 10 L^*$ for galaxies at a similar redshift. 

\begin{figure*}
    \centering
    \includegraphics[width=8.0 in]{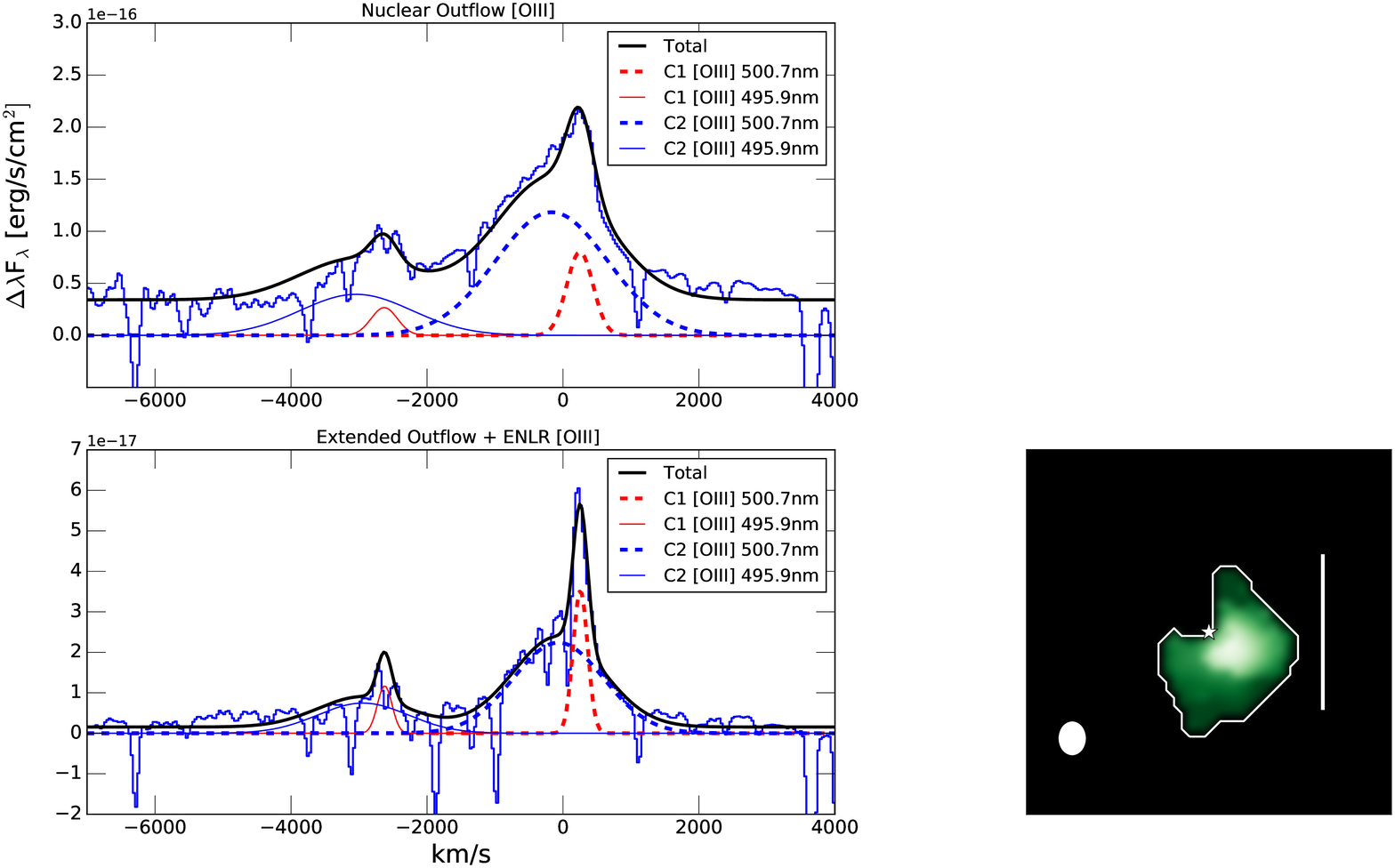}
    \caption{Top left: integrated spectrum from point source emission in the SDSSJ1652+1728 system along with multi-Gaussian fit to the \oiii\ emission line. Bottom left: integrated spectrum over the extended emission along with the fit model. On the bottom right in white contours we show the region that was integrated to produce the spectrum. The ellipse represents the size of the PSF and the bar represents 1 arcsecond on sky or about 8 kpc at the redshift of the target.}
    \label{fig:spec_SDSS1652}
\end{figure*}

\begin{figure*}
    \centering
    \includegraphics[width=8.0 in]{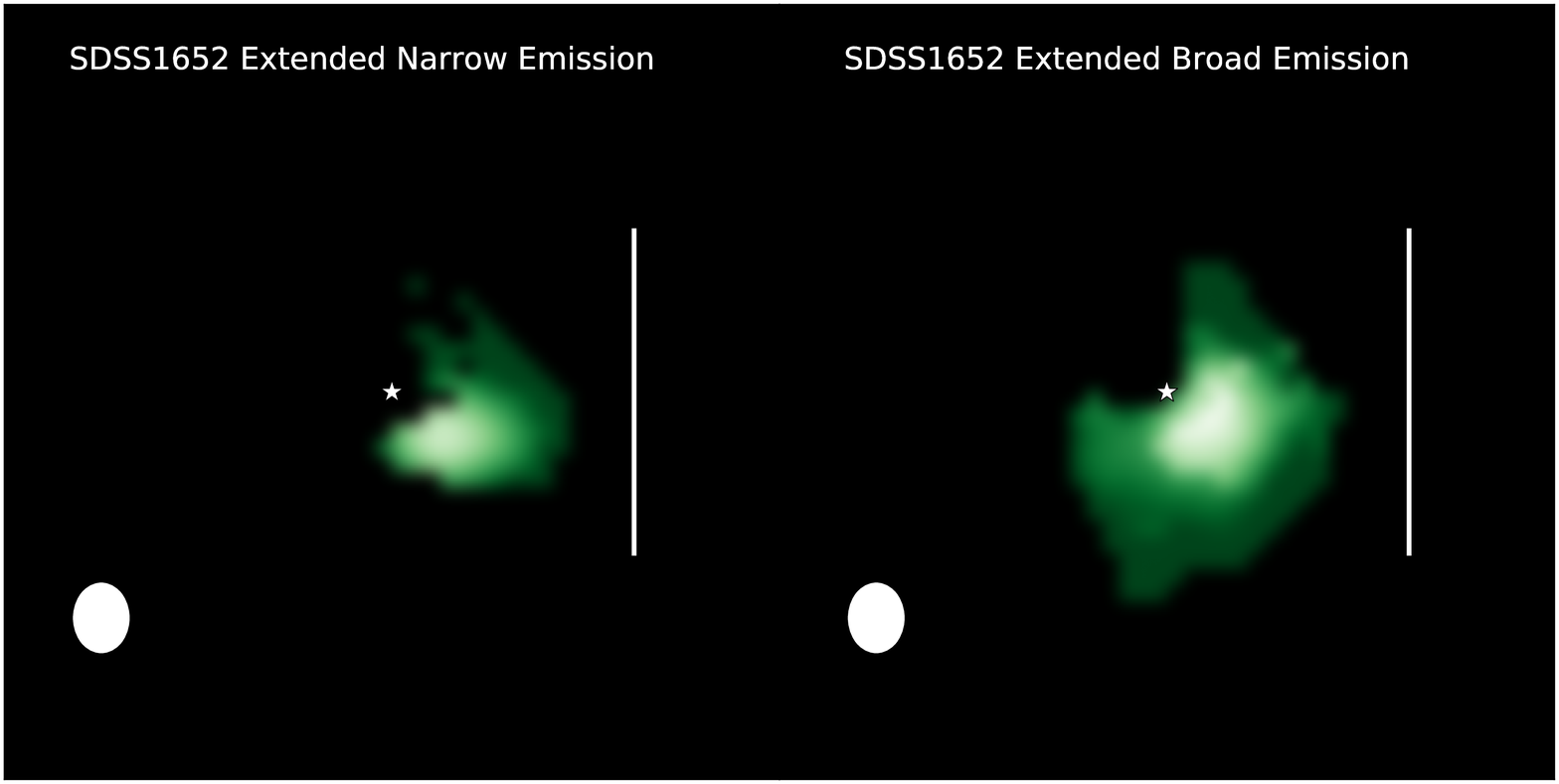}
    \caption{Figure comparing the extended narrow and broad emission morphology and distribution in SDSSJ1652+1728. Left: Extended narrow \oiii\ emission line map. Right: Extended broad \oiii\ emission line map. The ellipse in the lower left corner shows the FWHM of the PSF while the bar represents 1\arcsec.}
    \label{fig:SDSS1652_broad_n_narrow}
\end{figure*}

After performing PSF-subtraction on the IFS observations, we detected extended emission towards the south-west, consisting of both narrow and a broad emission line (Figure \ref{fig:1652}). The spectra along with the fits are shown in Figure \ref{fig:spec_SDSS1652}. We compare the distribution and morphology of the extended narrow and broad emission in Figure \ref{fig:SDSS1652_broad_n_narrow} by constructing \oiii\ emission line maps for each component that we fit in Figure \ref{fig:spec_SDSS1652}. The broader emission is associated with the outflow and is more extended towards the south, while the narrow emission extending towards the west is either gas within the host galaxy or a tidal tail feature associated with the extended emission seen in the HST observations. 

After subtracting a model of the extended emission, we detect unresolved emission on scales $<1.7$ kpc. We extract the spectrum from the point source. The spectrum consists of a broad \oiii\ emission line from the outflow and a narrow component, either associated with the host galaxy or a merger component. We detect a slight shift between the average velocity of the unresolved and resolved outflow. The unresolved component must be greater in size than the broad-line region because we are detecting it in a forbidden line. For example, the \oiii\ line is suppressed by collisional de-ionization at $n>8\times 10^5$ cm$^{-3}$, hence it must arise on scales of at least a few hundred pc \citep{hama11}. The majority of the emission in this object is from the unresolved component. The profile of the \oiii\ emission line is consistent with that is found in \citet{perr19}. Unfortunately, we cannot measure any interesting line ratios for this object because of its redshift, such that H$\beta$ falls right between $H$ and $K$ bands.

SDSS~J1652+1728 is part of an approved early-release-science (ERS) program for the \textit{James Webb Space Telescope (JWST)}. The observations will provide resolved \hb\ and \ha\ maps allowing us to conduct resolved photoionization diagnostics for this object. Furthermore additional access to rest frame near-infrared H$_{2}$ emission lines will allow us to constrain the multi-phase nature of the outflow in this system.

\subsection{SDSS~J0812+1819 (z=2.3794)}
\label{sec:0812}

This object was identified as a type-2 quasar candidate by \citet{alex13} based on its rest-frame UV properties. Follow-up NIR (rest-frame optical) spectroscopy by \citet{gree14b} did not reveal a broad component in H$\beta$ but did reveal one in H$\alpha$; hence this source can be classified as type 1.9 according to the standard optical line classification scheme. The object matches to an extended radio source in the FIRST survey \citep{beck95} consisting of two resolved radio sources centered on the quasar offset by 3$-$5\arcsec\ from the optical position and with an integrated flux density of 5.5 mJy. With this flux and the observed narrow \oiii, this object's radio emission is not consistent with being wind-induced \citep{zaka14, hwan18}, and we classify it as radio-loud. The morphology is linear and consistent with a lobe-dominated source at position angle 170$^{\circ}$ East of North. Further, 33 GHz observations reveal a core with flux 0.23 mJy/beam and possible traces of optically-thin emission from the lobes \citep{alex17}. 

\begin{figure*}
    \centering
    \includegraphics[width=8.0 in]{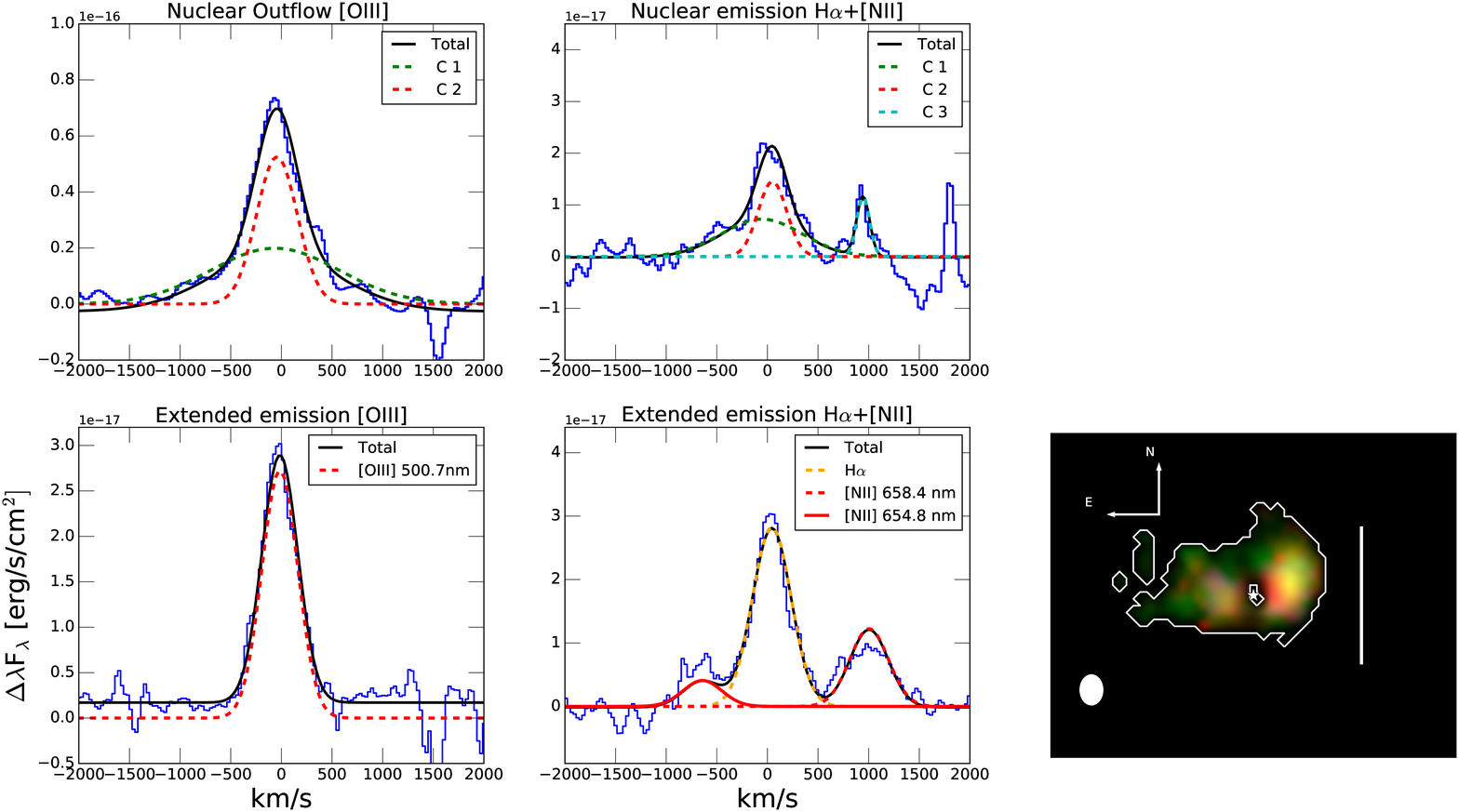}
    \caption{Top left: Spectrum from point source emission in the SDSSJ0812+1819 system along with multi-Gaussian fit to the \oiii, \ha\ and \nii\ emission lines, the broad-\ha\ component has been subtracted out to highlight the outflow in \ha. Bottom left: integrated spectrum over the extended emission along with the best fit model. On the bottom right in white contours we show the region that was integrated to produce the spectrum. The ellipse represents the size of the PSF and the bar represents 1 arcsecond on sky or about 8 kpc at the redshift of the target.}
    \label{fig:spec_SDSS0812}
\end{figure*}

After performing PSF subtraction on the IFS observations, we detect extended emission associated with two clumpy regions extended in the east-west direction in the \oiii, \ha, and the \nii\ emission lines. We detect no outflows on these spatial scales. After subtracting a model of the extended emission (similar to SDSSJ1652+1728, section \ref{sec:1652}), we detect an unresolved component that consists of both narrow and broad emission components. The narrow emission comes from the extended-narrow line region of the quasar, while the broad component is associated with an outflow on scales $<1.9$ kpc. The spectra along with the fits are shown in Figure \ref{fig:spec_SDSS0812}. 

Given that we detect \oiii, \ha, and the \nii\ emission lines in this object, we can use line ratios to investigate the source of ionization across the galaxy. Since we do not detect \hb\ in individuals spaxles, we calculate its flux assuming case B recombination ($\rm F_{H\alpha}/F_{H\beta}$ = 2.87) and no dust extinction. Under these assumptions, we find that 73$\%$ of the spaxels are consistent with gas photoionization from massive stars as they fall within the maximum expected line ratios for star formation on the BPT diagram at $z=2$ \citep{Kewley13b}. The rest of the spaxels are consistent with quasar photoionization. We combine the \ha\ flux from spaxels consistent with star formation, and we measure a star formation rate of 46$\pm$5 \myr\ using the empirical \ha\ luminosity star formation rate relation form \citep{kenn98}. Likely, extinction is contributing to a lower \logohb since we are using \ha\ and case B recombination as a proxy for \hb. More accurate extinction values are necessary to measure the actual fraction of the gas photoionized by stars vs. the quasar. Integrating over the extent depicted in contours shown in Figure \ref{fig:spec_SDSS0812} we measure a \loghn ratio of $-0.49$ and a \logohb value of 0.65; however, the actual value is likely closer to 1 based on the \hb\ flux limit from the total integrated spectrum in Figure \ref{fig:spec_SDSS0812}.

\subsection{SDSS~J2323$-$0100 (z=2.3695)}
\label{sec:2323}

Like SDSS~J1652+1728, SDSS~J2323$-$0100 was selected as an ERQ \citep{ross15, hama17}. A noteworthy feature of its rest-frame UV spectrum from the SDSS \citep{zaka16b} is a high apparent \nv\ 1240 \AA/Ly$\alpha$ ratio, even higher than that seen in SDSS~J1652+1728 \citep{alex18}; this feature may be due to outflows in the broad-line region wherein the high-velocity \nv\ troughs may be blocking Ly$\alpha$ \citep{hama17}. SDSS~J2323$-$0100 was among the first objects we observed with NIR (rest-frame optical) spectroscopy, revealing some of the most extreme kinematics of the \oiii\ emission known, with FWHM=2625 \kms\ \citep{zaka16b, perr19}. HST observations of this object reveal a face-on major merger between two galaxies, with nuclei separated by 1.3\arcsec=11 kpc and the SDSS position of the quasar consistent with the south-western of the two galactic nuclei. Due to the lack of previous integral-field or spatially resolved spectroscopy, it was unknown whether the second nucleus hosts any activity. The source is not detected in the radio down to 0.03 mJy and is classified as radio-quiet \citep{hwan18}. 

\begin{figure*}
    \centering
    \includegraphics[width=7.5 in]{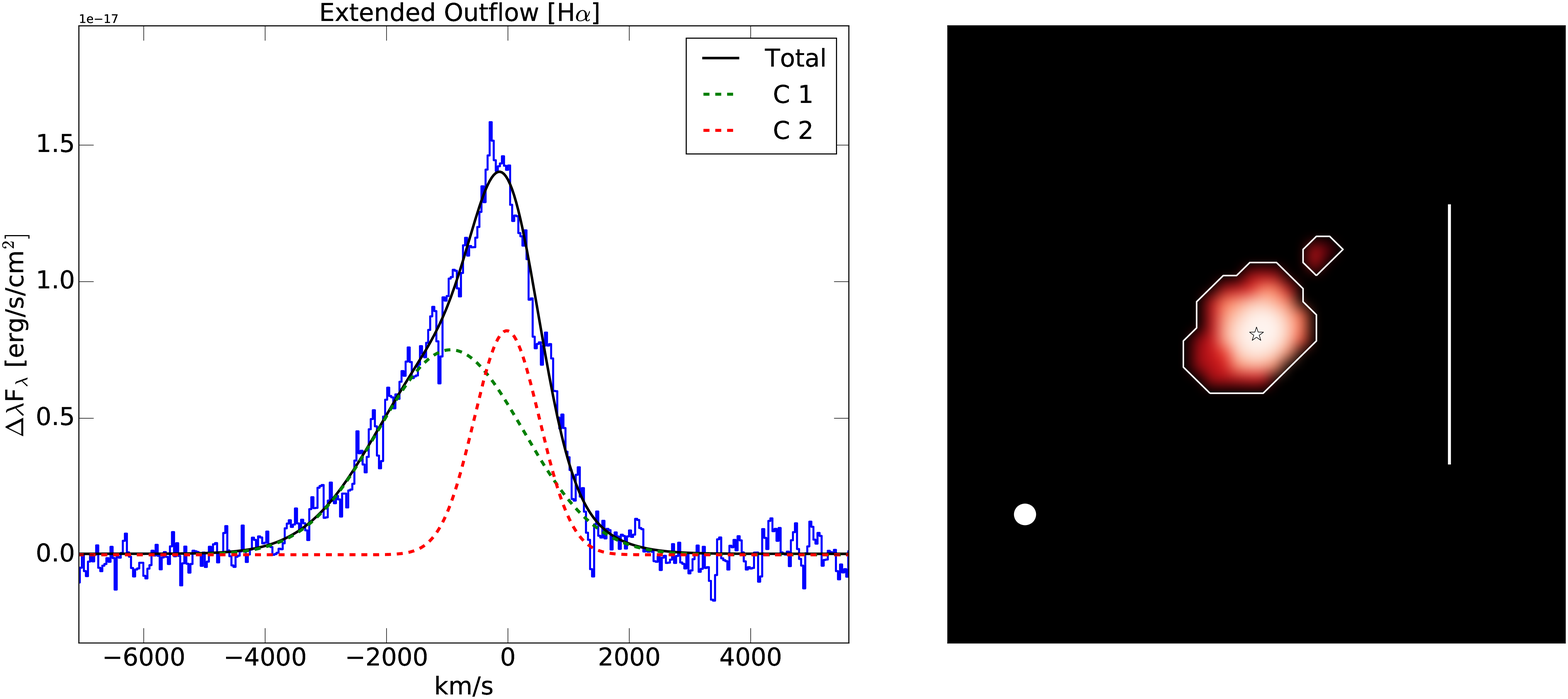}
    \caption{Left: Spectrum from extended emission in the SDSSJ2323-0100 system along with multi-Gaussian fit to the \ha\ emission line. Right: in white contours, we show the region that was integrated to produce the spectrum. The ellipse represents the PSF size, and the bar represents 1 arcsecond or about 8.6 kpc at the redshift of the target.}
    \label{fig:spec_SDSS2323}
\end{figure*}

\begin{figure}
    \centering
    \includegraphics[width=3.7 in]{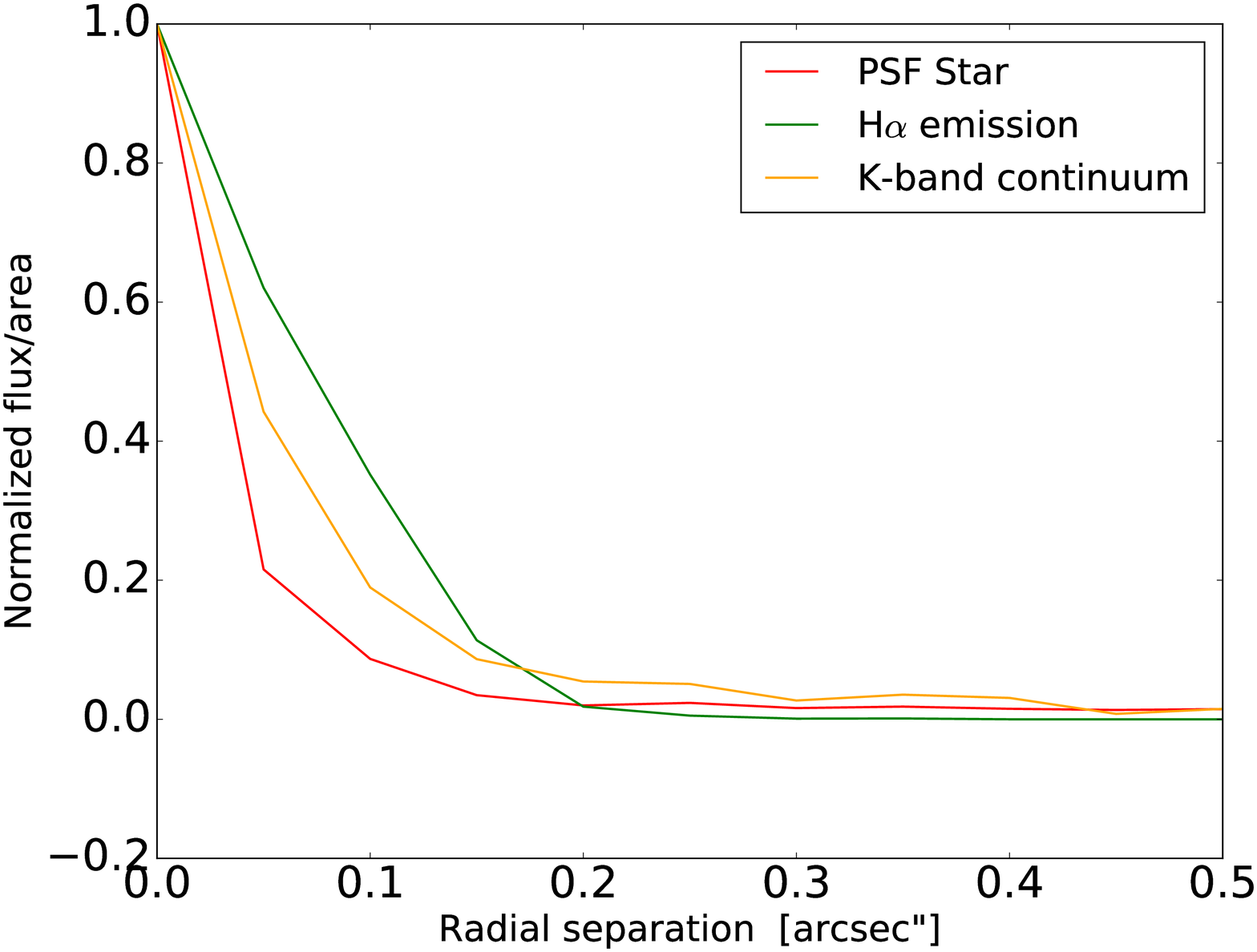}
    \caption{Radial profile analysis for SDSSJ2323-0100 to gauge the extent of the continuum and line emission. The curves represent azimuthally averaged radial profiles for the PSF star (red), \ha\ emission map (green) and the K-band line free continuum (orange).}
    \label{fig:radial_profile_SDSS2323}
\end{figure}

PSF-subtracted OSIRIS observations reveal a marginally spatially resolved emission in \oiii\ and \ha\ associated with the galaxy hosting the luminous quasar. The \oiii\ emission line profile is consistent with a single broad component, while the \ha\ emission line has both a broad and and an intermediate velocity (1267$\pm$54 \kms) component. The difference between the \ha\ and \oiii\ profiles within our observations likely due to insufficient wavelength coverage near the \oiii\ emission lines to perform accurate emission-line and continuum fitting. The spectra along with the fits are shown in Figure \ref{fig:spec_SDSS2323}. We have also attempted to search for evidence of rest-frame \nii\ 6549 \AA\ and 6584 \AA\ emission lines, by including additional Gaussians in the fit. When fitting for the \nii\ emission lines the ratio between \nii\ 6584 and \nii\ 6549 were held fixed at 1:2.95, with the lines width and redshift tied to \ha, hence the only free parameter was the flux of the \nii\ 6584 emission line. The resultant fit had \nii\ emission line flux within the noise of our observations and resulted in a poorer $\chi^{2}$ value. Likely, the \nii\ fit is below the threshold of our observations. The profile of the \ha\ emission line is consistent with that of the \oiii\ emission measured by \citet{perr19}. The continuum in the OSIRIS observations is also resolved on a similar scale as the emission lines and is consistent with stellar light from the host galaxy. No continuum is detected from the quasar, only emission from the broad-line region in \ha. The \textit{H}-band observations were conducted first and hinted at extended emission both in continuum and line emission when comparing their spatial profiles to that of the tip/tilt star. To perform a more accurate measurement of the size of the line and continuum emission, the $K$-band observations for this object were accompanied by observations of a nearby star at similar separation from the tip/tilt star as the quasar. We construct azimuthally average 1D profiles for the \ha\ emission and the continuum and compare them to the profile of the PSF-star, the tip/tilt star, and a type-1 quasar observed on the same night under similar conditions. The 1D profile of \ha\ and continuum emission is broader than that of the PSF-star consistent with being extended (Figure \ref{fig:radial_profile_SDSS2323}). We fit the PSF star with a 2D Gaussian profile and measure a FWHM of 0.082$\times$0.08\arcsec\, which we believe is an accurate representation of the resolution achieved on our observations. We deconvolve this value from the spatial profiles of \ha\ and continuum and measure spatial extents of $92\pm1$ mas for \ha\ at 90\% integrated light and an effective radius for the stellar continuum of 50$\pm$1 mas. We measure an emission line free $K$-band magnitude for the host galaxy of 20.7$\pm$0.1. Only continuum emission is detected in the nucleus of the merging galaxy and is likely consistent with stellar light, however it is too faint and falls close to the edge of the OSIRIS FOV to place an accurate magnitude.

While the OSIRIS observations reveal continuum emission, we are likely only detecting the circumnuclear emission from the galaxy. The 1-$\sigma$ surface brightness limit of our observations is 20.42 mag/\arcsec$^2$ and the rest of the structure detected in this galaxy with HST \citep{zaka19} is at a fainter surface brightness.

\subsection{SDSS~J0826+0542 (z=2.5767)}
\label{sec:0826}

SDSS~J0826$+$0542 was selected from the SDSS as an ERQ \citep{ross15, hama17}, and just like for SDSS~J1652$+$1728, and SDSS~J2323$-$0100, follow-up NIR (rest-frame optical) observations obtained with Gemini GNIRS revealed high-velocity, broad and blue-shifted component in the \oiii\ emission line \citep{goul18a, perr19}. It is an extremely luminous, hard-spectrum X-ray source consistent with near-Compton-thick absorption along the line of sight \citep{goul18a}. Just like SDSS~J1652+1728, this source is marginally radio-loud \citep{hwan18}.

\begin{figure}
    \centering
    \includegraphics[width=3.7in]{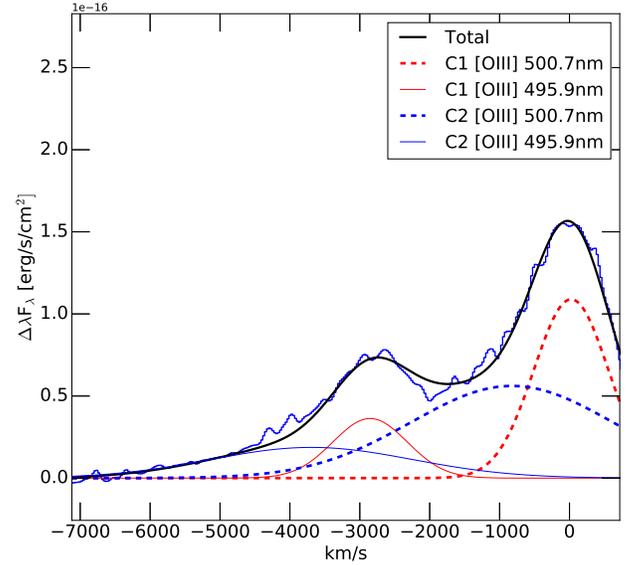}
    \caption{Spectrum from point source emission in the SDSSJ0826+0542 system along with multi-Gaussian fit to the \oiii\ emission lines.}
    \label{fig:spec_SDSS0826}
\end{figure}

After performing PSF subtraction we detect no extended emission. We extract the spectrum from the point source. The spectrum is consistent with a broad \oiii\ emission line coming from an outflow region. The upper limit on the size of the outflow is 0.4\arcsec. We present the spectrum in Figure \ref{fig:spec_SDSS0826}.

\section{Discussion}
\label{sec:disc}

\subsection{Comparison with other high redshift samples}

Over the last decade, numerous studies have focused on resolving ionized outflows in high redshift galaxies with AGN. These studies have utilized IFS \citep{harr12,cano12,carn16,vayner17} and long-slit spectroscopy \citep{Coil15,Leung19} targeting nebular emission lines redshifted into the near-infrared. Generally, they have been successful at detecting ionized outflows and measuring their dynamics and kinematics to better understand their impact on their respective galaxies. These studies, for the most part, used slightly different ways of calculating the mass outflow rates. These outflow rates have to be recomputed following the methodology presented in \citet{fior17} and \citet{vayner20} to consistently compare the results from different surveys.

\begin{figure*}
    \centering
    \includegraphics[width=7.5in]{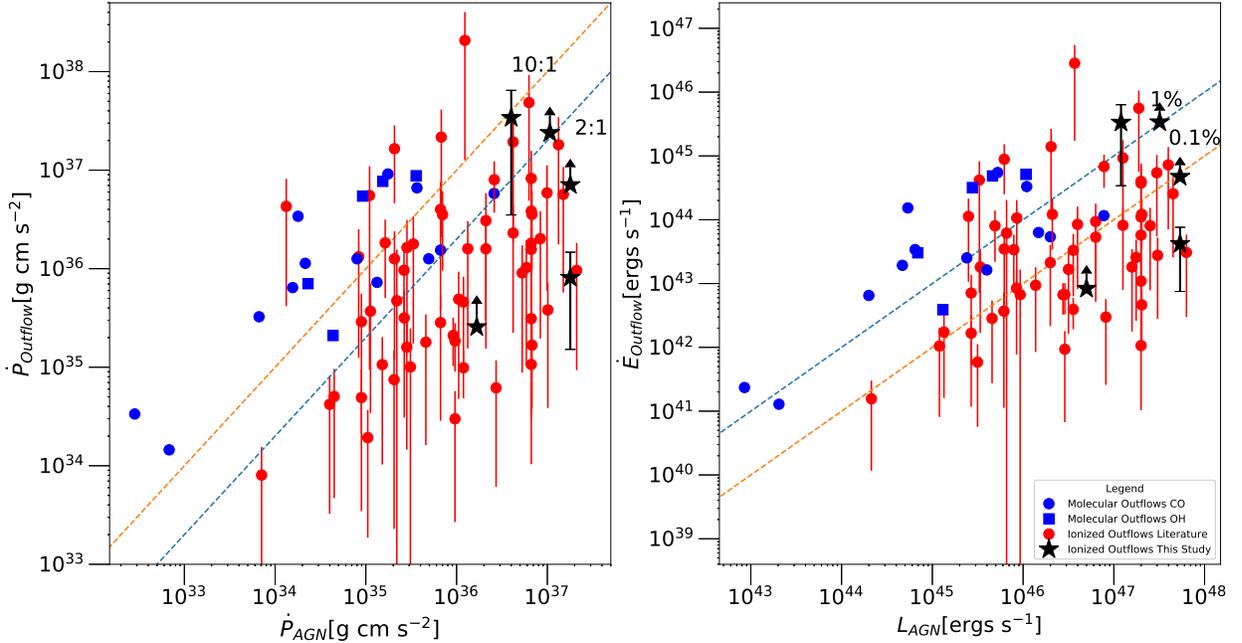}
    \caption{On the left we plot the momentum flux of the outflow against the radiation momentum flux of the accretion disk. Red points represent ionized outflows in the distant Universe, with the black stars representing the objects within our sample. The blue points are molecular outflows studied in low redshift galaxies. We plot two lines of constant ratio; 2:1 and 10:1 as these provide clues into the driving mechanism of the outflows. Theoretically points below the 2:1 signal outflows that are driven by either radiation pressure or a radiative shock. Points above the 2:1 line signal outflows that are driven by either radiation pressure in a confined region within $<1$ kpc or an outflow driven by an adiabatic shock. On the right we plot the energy rate of the outflow against the bolometric luminosity of AGN. The lines of constant ratio signal coupling efficiencies proposed by various theoretical works necessary to clear the gas from the host galaxy through outflows and establish the local scaling relations such as \msigma.}
    \label{fig:E_dot_P_dot}
\end{figure*}

\cite{vayner20} collated data from studies of galaxies with outflows at a redshift of $\rm 1<z<3$, at the peak of the accretion rate density on SMBH \citep{Delvecchio14} and the star formation rate density of the Universe \citep{mada14} where galaxies and SMBH undergo the majority of their growth. The galaxies span a wide range of star formation rate, AGN bolometric luminosity (L$_{bol} = 10^{45-47}$\ergs) and nuclear obscuration (both type-1 and 2). In \cite{vayner20}, the outflow rates are recomputed, taking into account some of the significant uncertainties that scourge outflow rates and energetics in the distant Universe. The literature outflow rates in \cite{vayner20} are combined with a sample of radio-loud type-1 quasars observed with OSIRIS in the same study. We compute our outflows rates in the same manner (see section \ref{sec:outflow-rates-energetics}) and, for the first time, can compare information on resolved outflows in ERQs to other samples in the distant Universe. In Figure \ref{fig:E_dot_P_dot}, we present the location of the ERQs on the outflow momentum flux (\momfluxout) vs. AGN radiation momentum flux diagram and the equivalent energy rate of the outflow vs. the luminosity of the AGN diagram.

\begin{figure*}
    \centering
    \includegraphics[width=7.5in]{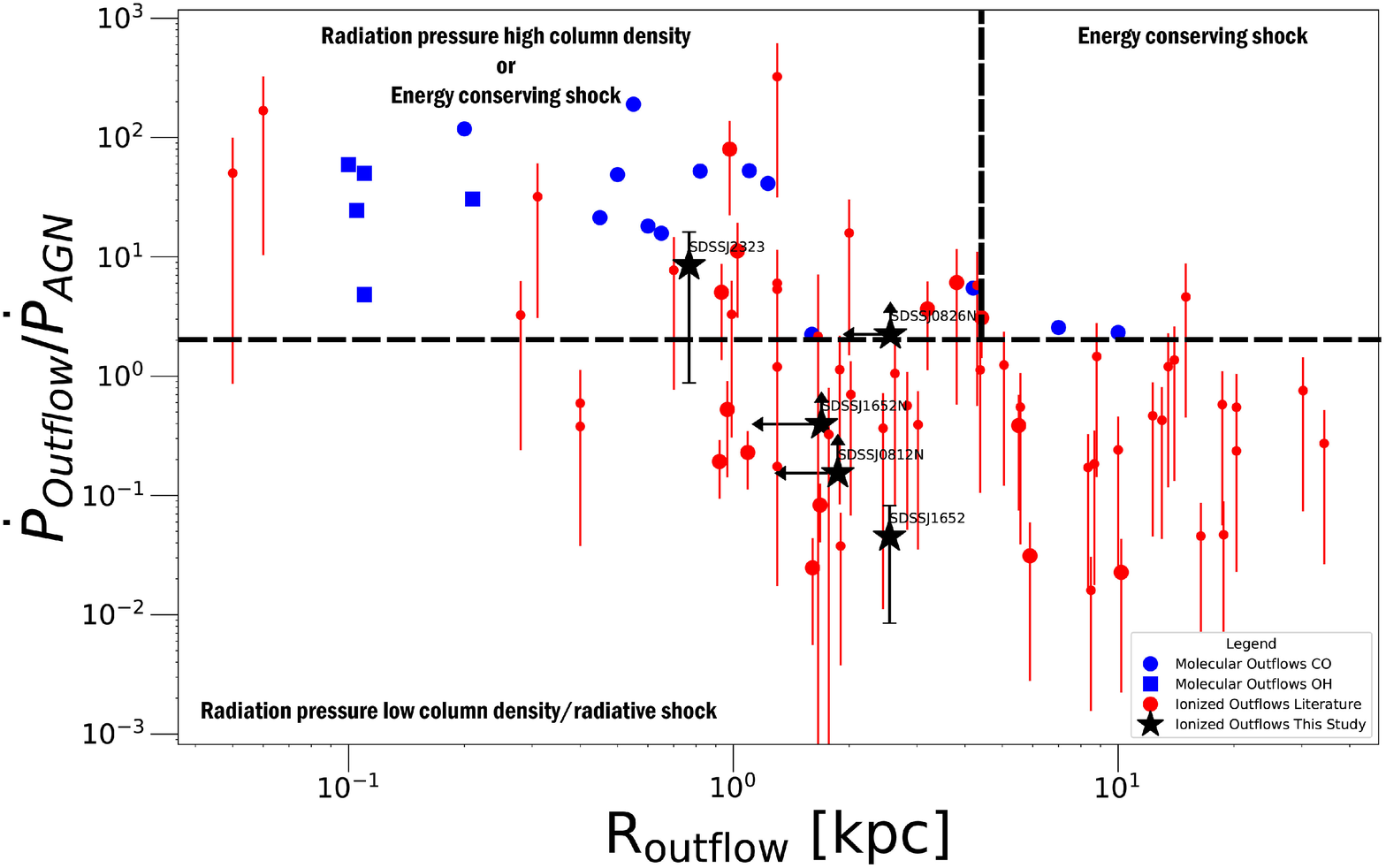}
    \caption{The ratio of momentum flux as a function of the outflow radius. Red points are from ionized outflows, and blue points are from molecular, the points from this study are shown with a black star. High momentum flux ratio on small scales is generally attributed to radiation pressure-driven winds in a high column density environment. High momentum flux on large and small scales can be attributed to an adiabatic shock driven outflow, while low momentum flux ratio on small and large scales can be attributed to a radiative shock. Low momentum flux ratios on small and large scales can also be attributed to radiation pressure-driven winds in low column density environments. The dividing lines between the models are selected from theoretical work by \citealt{fauc12a,thom15,costa18}.}
    \label{fig:P_dot_ratio_vs_R}
\end{figure*}

Similarly, we place the ERQs on the outflow momentum flux ratio vs. outflow radius plot shown in Figure \ref{fig:P_dot_ratio_vs_R}. The momenta and energy rates for ERQs are within the range of other high redshift studies at similar bolometric luminosity. However, interestingly the outflows within our sample all appear to be compact relative to the majority of the ionized outflows found in the distant Universe. Given our small sample, it is challenging to say anything about the general ERQ population and how that may differ from other studies.

\subsection{Outflow driving mechanisms}

Numerous mechanisms have been proposed to explain the physical mechanisms that drive galaxy scale outflows \citep{murr05,thom15,costa18,Veilleux20}. They include radiation pressure from AGN/massive young stars, supernova explosions, shocks driven by ultra-fast outflows, jets, and broad-absorption line winds. Often to distinguish between all driving mechanisms requires spatially resolved observations of outflows and measurement of their kinetic luminosity, momentum flux, and radial extent as well as energy and momentum supply from both star formation and AGN activity. Quasars tend to reside in galaxies with a broad range of star formation properties; understanding the star formation rates, as well as the AGN bolometric luminosities, therefore crucial \citep{Aird19}.

The ratio between the momentum flux of the outflow to the photon momentum flux of the accretion disk (L$_{AGN}$/c) or stellar bolometric luminosity and the radius of the outflow can be used to distinguish between various sources of driving mechanisms \citep{thom15,zubo12,costa18}. Current theoretical consensus is that high \momfluxratio\ on $>1$ kpc scales is due to adiabatic shocks, where the material is swept by an outflow that does not radiate energy efficiently. Generally, in these outflows, the momentum flux grows as the outflow expands until the outflow reaches a certain radius where gravitation effects and interactions with the ISM and circum-galactic medium slow the outflowing material \citep{fauc12b}. High \momfluxratio\ on $<$ 1 kpc scales can be attributed to a radiation pressure-driven wind in a high column density environment where infrared photons are trapped and scattered multiple times, which gets reflected in the ``momentum boost" \citep{thom15,costa18}. Typically radiation pressure-driven winds achieve a maximum \momfluxratio\ $\approx2$ on $>1$ kpc scale, with the \momfluxratio\ dropping down as the outflow expands. Low \momfluxratio\ $<$1 on any scale can be either attributed to radiation pressure-driven winds in a low column density environment or a shock that can radiate its energy efficiently \citep{fauc12b}.

For sources within our sample, multiple mechanisms are likely at play in driving the observed outflows. For SDSSJ2323-0100 and SDSSJ0826+0542, the potentially high momentum flux ratios tell us that an adiabatic shock driven by a ultra-fast outflow or a broad-absorption line type wind can drive the galaxy scale outflows. Given the relatively small radii, another likely scenario is the ionized outflows are driven by radiation pressure from the accretion disk of the quasars in a high column density environment. For SDSSJ1652+1728, radiation pressure from the quasar accretion disk or a radiative shock can drive the observed ionized outflow. At the extent of the outflow, there would have to be a factor 20 more outflowing gas mass in either the neutral or molecular phase for the outflow to be considered driven by an adiabatic shock or radiation pressure from a high column density environment \citep{costa18}. Other studies have found a substantial amount of neutral and molecular gas in galactic outflows \citep{carn15,vayner17,rupk17,Fluetsch19}. The factor of 20 is based on how much we would need to scale the \momfluxratio\ of the outflow in SDSSJ1652+1728 at its current extent to match that found in simulations by \citet{fauc12a} for an energy conserving outflow. The factor of 20 is also consistent with predictions by \cite{demp18} for the amount of neutral or molecular gas required in an outflow to prevent over-ionization of the \oiii. For SDSSJ0812+1819, the outflow is likely driven by either radiation pressure or a radiative shock. For an adiabatic shock to drive the outflow, there would have to be a factor of 40 or more gas in a neutral or molecular gas phase based on the momentum fluxes found in simulations by \citet{fauc12a} at the extent of the outflow in SDSSJ0812+1819. A nuclear starburst would require a star formation rate of 2300 \myr, 2000 \myr, 8,000 \myr, and 50 \myr\ for SDSS2323-0100, SDSS1652+1728, SDSS0826+0542 and SDSS0812+1819, respectively, to drive the outflow through radiation pressure. We assume the Starburst99 stellar feedback model \citep{leit99} with a Kroupa IMF \citep{kroup01}:

\begin{equation}
     \dot{P}_{SFR} = 1.5\times10^{33}\frac{\dot{M}_{SFR}}{1M_{\odot}yr^{-1}} \rm dyne\label{eq:pdot_SFR_SFR}
\end{equation}

For star formation to drive the outflows through supernova explosions would require a star formation rate of 500 \myr, 500 \myr\ and 1600 \myr\ for SDSS2323-0100, SDSS1652+1728, and SDSS0826+0542. We assume there is one supernova explosion for 100 solar mass of stars formed and using the momentum injection per supernova from recent simulations by \citet{Kim15} and \citet{Martizzi15}:

\begin{equation}
    \dot{P}_{SNe} = 7.5\times 10^{33} \frac{\dot{M}_{SFR}}{1M_{\odot}yr^{-1}} \left(\frac{n}{100\,{\rm cm}^{3}}\right)^{-0.18}
    \left(\frac{Z}{Z_\odot}\right)
    \rm dyne \label{eq:SNe_mom}.
\end{equation} 

Such high star formation rates are not supported by our data, as we see no evidence for narrow \ha\ or \hb\ emission that would be consistent with nuclear starbursts at such star formation rates. There could be obscured star-forming regions, and future observations with ALMA can help search for that. Most likely, the entirety of the outflows are driven by radiative mechanisms of the quasar. SDSSJ0812+1819 shows evidence for jets in radio data, hence there is a possibility of jets also driving the observed nuclear outflow in this one system. For SDSS1652+1728, given the redshift we cannot currently put a limit on the star formation rate using a Balmer emission line. The existence of a narrow \oiii\ component could signal photoionization by massive young stars in an HII region, however, just as likely, quasar photoionization can produce the observed emission line. Interestingly, the overall orientation of the ionized gas outflow relative to the nucleus is consistent with the 220 deg East of North (center of the SW quadrant) inferred from the polarimetric observations. The detection of co-spatial scattering and forbidden-line outflows indicates that the quasar illuminates both the outflowing gas and the scattering gas with the same geometry \citep{wyle16a}. 

\begin{figure*}
    \centering
    \includegraphics[width=7.0 in]{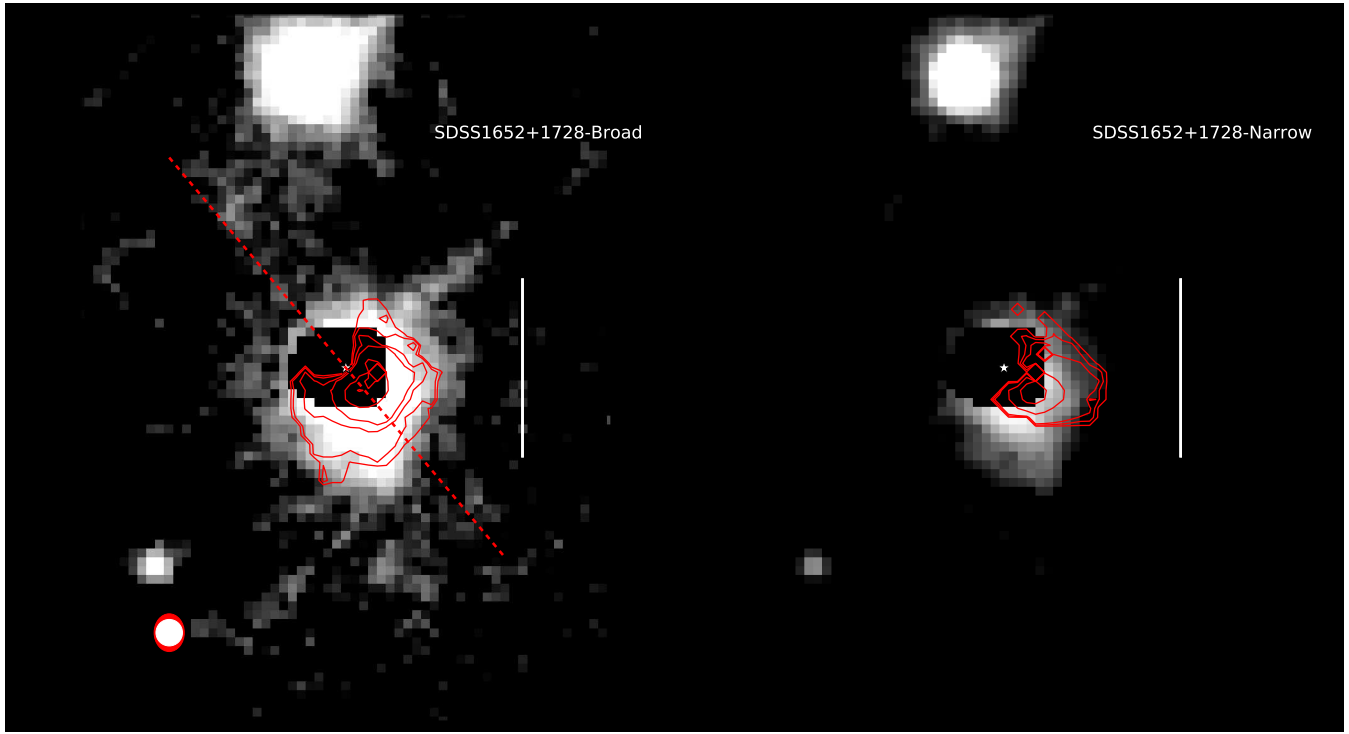}
    \caption{\textit{HST} ACS observations of SDSS1652+1728. In the background of both images, we show a PSF subtracted F814W image of the host galaxy. On the left, we overlay in contours the integrated intensity \oiii\ map from the broad component due to the outflow, and we stretch the ACS image in the range of 22.5 to 24.5 mag/arcsec$^{2}$ to highlight the faint diffuse emission, the dashed line represents the scattering angle from spectropolarimetric observations \citep{alex18}. On the right, we overlay the narrow component of \oiii, that originates either from star-formation or quasar photoionization from an extended narrow-line region. Only a small fraction of the UV emission correlates with the narrow \oiii\ component. The image stretch on the right ranges from 21.5 to 23.5 mag/arcsec$^{2}$ to highlight the brighter compact emission. The strong correlation between the extended broad \oiii\ and UV emission indicates that the majority of the extended UV light ($\lambda_{rest}$ = 2733.8 \AA) originates from scattering in the direction of the outflow. The bar in each image represents 8.5 kpc, the red ellipse in the lower left corner represents the resolution of the NIFS observations, while the white ellipse shows the resolution of \textit{HST} ACS.}
    \label{fig:ACS}
\end{figure*}

Further evidence for a co-spatial UV photon scattering and outflowing gas comes from our recent \textit{Hubble Space Telescope} ACS observations (GO-14608, PI: Zakamska, exposure time: 2188.0 s) taken in the F814W filter ($\lambda_{rest}$ = 2733.8 \AA). Spectropolarimetric observations already revealed a high degree of polarization near these wavelengths; after performing PSF subtraction of the quasar emission on the ACS data, we detect extended emission in the SW - NE direction, matching closely to the predicted orientation of illumination from spectropolarimetry \citep{alex18}. Within the wavelength range of the F814W filter, we do not expect any strong emission from UV lines; hence nearly the entire light in this filter has to come from a continuum source. The PSF subtraction was done by extracting a portion of the ACS image centered on a nearby star with a similar magnitude as the quasar; we then normalized the image to the peak flux value and re-scale it the peak flux of the quasar, and subtract it out. We compare the extent and morphology of the rest-frame UV emission to both the narrow and broad emission detected in \oiii. The general orientation, extent, and morphology of the rest-frame UV emission match closely to the extended broad-emission detected in \oiii\ from the outflow. The narrow \oiii, which extends in the western direction, matches a portion of the UV emission that likely originates from a star-forming region in either the host galaxy or a merging galaxy (Figure \ref{fig:ACS}). The direction of the narrow \oiii\ emission also matches the general extent of the WFC3 F140W imaging from \citep{zaka19}, that traces the stellar continuum from the host galaxy.\\ \indent Most of the UV emission likely comes from scattered light in the conical illumination region for SDSS1652+1728. The scattered light's extent is approximately two times greater than the ionized outflow detected with NIFS. The diameter of the extended UV emission is approximately 24 kpc. The \oiii\ emission appears to trace a single cone of the outflow, while in the UV light, we see a double-cone, indicating that the ionization may be impacting a much larger extent. The asymmetry of the UV light likely indicates that scattering is occurring of dust grains in the outflow. Forward scattering can explain the differences in the observed intensity between the SW and NE regions if the axis of the scattering cone points towards the observer in the SW region. Presence of dust can also explain why the NW cone is undetected in \oiii, where the back cone is partially extincted by dust and the \oiii\ line strength is below the flux sensitivity of NIFS. Based on \citet{zaka05,obie16} scattering models due to dust grains and electrons in z$\sim 0.5$ type-2 quasars, the scattered light comes from low-density clouds. For large scattering regions, dust scattering is further favored over electron scattering since, based on the modeled electron density, the recombination emission line intensity would be unphysical \citep{zaka05}. The low density of these regions can partially explain the larger extent of the scattering light compared to \oiii, due to the stronger dependence of the \oiii\ intensity on the electron density. The direction of quasar illumination and direction of scattered UV photons matching closely to the galactic outflow further support that the quasar is the primary driving source of the galactic outflow. Accurate measurement of the energetics from the neutral and molecular phase is imperative to understand both the driving mechanism behind the galactic scale outflow and the impact on the star formation in the host galaxy.

\subsection{Potential impact of the outflow on their host galaxies}

Several theoretical works have proposed a coupling efficiency between the galaxy scale outflow and the quasar's bolometric luminosity to explain the observed local scaling relation between the mass of the SMBH and the velocity dispersion and bulge of a galaxy. Generally, in these theoretical frameworks, once the galaxies assemble onto a local scaling relation, a certain fraction of the quasar bolometric luminosity needs to be converted into the kinetic luminosity of the outflow to affect the flow of gas onto the galaxy, induce turbulence in the ISM to affect the efficiency of star formation \citep{power11,zubo12}. The current consensus is that if the quasars transfer about 0.1-1$\%$ of its luminosity into an outflow, there can be a significant impact \citep{Hopkins10,choi12,costa18}. We compute and present the ratio between the bolometric luminosity of the quasar and the kinetic luminosity of the outflow in Table \ref{tab:outflow-prop}.

For SDSSJ2323-0100 and SDSSJ0826+0542 we find that 2.7\% (0.2-5.2\%) and 1\% (0.5-1.5\%) of the quasars' bolometric luminosity is converted into the outflows' kinetic luminosity, respectively. Even at the low end of both ranges, the coupling efficiency is large enough for the outflow to affect the galaxy. For SDSSJ1652+1728, the coupling efficiency is 0.1\% (0.05-0.2\%), on the low end of what is prescribed by theoretical work. The addition of other ionization states of Oxygen and additional neutral and molecular gas components will increase the coupling efficiency. Furthermore, we assumed solar metallicity when computing the outflow mass using the \oiii\ emission line; if the true metallicity is lower, that can increase the amount of ionized gas that we compute. Hence the outflow in this system is likely to impact the galaxy. For SDSSJ0812+1819, the coupling efficiency is at 0.02$\%$, significantly lower than what is necessary for the outflow to impact the host galaxy, a substantial fraction of outflowing gas in either neutral or molecular phase would be necessary.

From a theoretical standpoint, the nuclei transition to an unobscured phase once a powerful nuclear outflow can be driven \citep{hopkins05,king15} efficiently. In the current theoretical paradigm, this is supposed to occur when the galaxies fall on the local \msigma relation because star formation in the host galaxy needs to be suppressed on a similar timescale to the SMBH growth. Driving outflows through radiation pressure or an adiabatic shock are most efficient in confined circumnuclear environments, producing a high \momfluxratio fast-moving turbulent outflows \citep{costa18,fauc12a,king15}. These conditions are potentially best satisfied in systems with high nuclear column densities and already known fast-moving outflows, making ERQs ideal candidates for long-sought transitional objects. In our two resolved objects, the outflow sizes are smaller than the average size measured in other AGN systems at similar redshift and bolometric luminosity. The outflows in SDSSJ2323-0100 and SDSSJ0826+0542 carry a significant amount of energy and momentum with a large momentum flux ratio that can be caused by radiation pressure or an adiabatic shock. The outflows are powerful enough to clear the gas and dust in the circum-nuclear region and transition these object to a blue type-1 quasar. For the above reasons, SDSSJ2323-0100 and and SDSSJ0826+0542 excellent example of transition objects in the distant Universe. 

While there is evidence for strong outflows in the SDSSJ2323-0100, SDSSJ1652+1728 and SDSSJ0826+0542 systems, we still find the presence of quiescent gas due to the narrow line emission in SDSSJ1652+1728 and the existence of a large molecular reservoir in the SDSSJ2323-0100 system (Serena Perrotta, private communication). Likely the outflows only affect the nuclear regions of these galaxies as they have yet to expand deeper into the interstellar medium. There is likely to be a delay between the present time and when the galaxy's full extent will be affected by these outflows. In SDSSJ0812+1819, the outflow is likely not powerful enough to affect any region of its host galaxy.

\section{Conclusions}
\label{sec:conc}
Extremely red quasars at high redshifts display some of the most extreme kinematics in their forbidden emission lines, and it is very important to determine whether the observed high velocities of the gas imply large-scale outflows and powerful energetics of feedback. In this paper, we present the first adaptive optics assisted NIR IFS (rest-frame optical) observations of 3 ERQs and a candidate type-2 quasar. The observations were conducted with the NIFS instrument at the Gemini North Observatory and with OSIRIS at the W. M. Keck Observatory.

We performed PSF subtraction to remove any unresolved quasar emission to search for extended emission from the host galaxies. We fit the emission lines at each spaxel to extract information on the line intensity, radial velocity offset, and dispersion. We detected extended emission in SDSSJ2323-0100 and SDSSJ1652+1728 originating from both a narrow-line region and an outflow, with only narrow emission detected in SDSSJ0812+1819. In SDSSJ0826+0542, no extended emission is detected, and we place an upper limit on its size. Outflows are found in all objects. In SDSSJ1652+1728, we detect both a resolved and an unresolved component to the outflow. In SDSSJ2323-0100, we only detect an extended outflow while in SDSSJ0812+1819 and SDSSJ0826+0542, the outflows are unresolved within our observations. We calculated outflow rates, kinetic luminosities, and momentum fluxes for all outflows. We compared the energetics to theoretical predictions and other high redshift samples and found:

\begin{itemize}

\item Several AGN mechanisms can drive the observed outflows. For SDSSJ2323-0100, either radiation pressure from the quasar or an adiabatic shock sweeping material in the host galaxy is responsible for driving the galaxy scale ionized outflow. For SDSSJ1652+1728, the outflow can potentially be driven by radiation pressure or a radiative shock, however outflowing gas in other phases can shift the driving mechanism to an adiabatic shock or radiation pressure from a high column density nuclear region. The outflow in SDSSJ0812+1819 is likely driven by radiation pressure in a low column density environment or a radiative shock.

\item The outflows in SDSSJ2323-0100, SDSSJ1652+1728, and SDSSJ0826+0542 satisfy the minimum coupling criteria between the bolometric luminosity of the quasar and kinetic luminosity of the outflow to have an impact on the star-forming properties of the host galaxy. However, there is evidence for narrow emission in SDSSJ1652+1728 indicating the presence of a quiescent gas while in SDSSJ2323-0100, there is a substantial molecular gas reservoir. Likely the outflows are only affecting at present the nuclear regions of these two galaxies. The coupling efficiency in SDSSJ0812+1819 is below the minimum value prescribed by theoretical work for the outflow to have any impact on the host galaxy.

\item We find the ionized outflows in our sample of ERQs to be less extended than ionized outflows in other samples in the distant Universe. The small outflow radius, high momentum flux ratio, and the high obscuration in SDSSJ2323-0100 make this an excellent candidate for a transitional object where the quasar driven outflow is expelling material from the nuclear region and unveiling the SMBH accretion disk.

\end{itemize}

\section{Data availability}
The Keck OSIRIS data of this work are publicly available from the Keck Observatory Archive (https://www2.keck.hawaii.edu/koa/public/koa.php). The Gemini NIFS data of this work are publicly available from the Gemini Observatory Archive (https://archive.gemini.edu/searchform). Source information is provided with this paper. Other data underlying this article will be shared on a reasonable request to the corresponding author.

\pagebreak
\section{Acknowledgments}
Based on observations obtained at the Gemini Observatory, which is operated by the Association of Universities for Research in Astronomy, Inc., under a cooperative agreement with the NSF on behalf of the Gemini partnership: the National Science Foundation (United States), the National Research Council (Canada), CONICYT (Chile), the Australian Research Council (Australia), Minist\'{e}rio da Ci\^{e}ncia, Tecnologiae Inova\c{c}\~{a}o (Brazil) and Ministerio de Ciencia, Tecnolog\'{i}a e Innovaci\'{o}n Productiva (Argentina). The authors thank Gemini scientists Julia Scharw\"achter, Trent Dupuy, Andreea Petric, Marie Lemoine-Busserolle, and Susan Ridgway for helping us plan and execute this program and who served as our Gemini phase II liaison for planning this program (GN-2019A-FT-111, GN-2016A-FT-8, GN-2014A-Q-42, GN-2014B-Q-44). Based on observations with the NASA/ESA Hubble Space Telescope
obtained at the Space Telescope Science Institute, which is operated by the Association of Universities for Research in Astronomy, Incorporated, under NASA contract NAS5-26555. Support for program number HST-GO-14608 was provided through a grant from the STScI under NASA contract NAS5-26555. The authors wish to thanks Jim Lyke, Randy Campbell, and other SAs with their assistance at the telescope to acquire the Keck OSIRIS data sets. The data presented herein were obtained at the W.M. Keck Observatory, which is operated as a scientific partnership among the California Institute of Technology, the University of California and the National Aeronautics and Space Administration. The Observatory was made possible by the generous financial support of the W.M. Keck Foundation. The authors wish to recognize and acknowledge the very significant cultural role and reverence that the summit of Maunakea has always had within the indigenous Hawaiian community. We are most fortunate to have the opportunity to conduct observations from this mountain. This research has made use of the NASA/IPAC Extragalactic Database (NED) which is operated by the Jet Propulsion Laboratory, California Institute of Technology, under contract with the National Aeronautics and Space Administration. A.V. and N.L.Z. acknowledge support of the STScI grant JWST-ERS-01335.002 and HST-GO-14608. R.A.R thanks partial financial support from  the Funda\c c\~ao de Amparo \`a pesquisa do Estado do Rio Grande do Sul (17/2551-0001144-9 \& 16/2551-0000251-7) and Conselho Nacional de Desenvolvimento Cient\'ifico e Tecnol\'ogico (202582/2018-3 \& 302280/2019-7). We want to thank the anonymous referee for their constructive comments that helped improve the manuscript.

\bibliography{master2}{}
\bibliographystyle{aasjournal}

\end{document}